\documentclass[conference, twocolumn]{IEEEtran}

\usepackage{graphicx}      

\usepackage{amsmath,amsfonts,amssymb}
\usepackage{cite}

\usepackage[latin1]{inputenc}
\usepackage[T1]{fontenc}

\usepackage{mathptmx} 

\newcommand{\taus}{\ensuremath{\tau_\sigma}}

\newcommand{\Fourier}{\ensuremath{\mathcal F}}

\begin{document}

\title{Four ways to justify temporal memory operators in the lossy wave equation} 

\author{Sverre Holm,\\ Department of Informatics, University of Oslo,\\P. O. Box 1080, Blindern, N-0316 Oslo, Norway}

\maketitle

\begin{abstract}                
Attenuation of ultrasound often follows near power laws which cannot be modeled with conventional viscous or relaxation wave equations. The same is often the case for shear wave propagation in tissue also.  

More general temporal memory operators in the wave equation can describe such behavior. They can be justified in four ways: 1) Power laws for attenuation with exponents other than two correspond to the use of convolution operators with a temporal memory kernel which is a power law in time. 2) The corresponding constitutive equation is also a convolution, often with a temporal power law function. 3) It is also equivalent to an infinite set of relaxation processes which can be formulated via the complex compressibility. 4) The constitutive equation can also be expressed as an infinite sum of higher order derivatives. 

An extension to longitudinal waves in a nonlinear medium is also provided.


\end{abstract}


\section{Introduction}

Arbitrary power law attenuation and dispersion in medical ultrasound and elastography can be modeled with temporal memory operators in the wave equation. Early modeling of such attenuation was \textit{inductive} because the wave equations were designed only to fit experimental measurements \cite{Szabo94,Chen04}. 

As a step towards a more \textit{deductive} approach, wave equations were derived from modified constitutive equations in combination with conservation laws, see e.g. \cite{holm2010} building on \cite{Caputo1967}. In this way, satisfaction of causality and the Kramers-Kronig conditions is ensured. This approach can be improved upon even  more if the constitutive equations can be found from even more fundamental physical mechanisms.

The purpose of this paper is to summarize the principles that power law wave equations and modified constitutive relations build on. Four alternative formulations will be given:
\begin{enumerate}
\item If the measured attenuation follows a power law, the memory operator can be shown to be a convolution with a temporal power law. This can also be expressed as a fractional derivative, opening up for a host of useful properties from mathematics. 

\item These wave equations can be derived from constitutive equations that include a convolution with a temporal power law. For some materials, such laws can be verified by time domain measurements.

\item A common method for simulation of power law attenuation is to use a few relaxation processes. This is not as ad hoc as it may look, as it can be shown that a temporal power law loss operator is equivalent to an infinite sum of exponential time domain operators. Simulation with a few such processes is therefore equivalent to a truncation of  a sum of relaxation processes.

\item The equivalence with a sum of relaxation processes opens up for modeling the constitutive equation alternatively with higher order derivatives. This means that a sum of second, third, and higher order derivatives may describe the fractional constitutive equation.

\end{enumerate}

All of these four descriptions are equivalent and bring new perspectives to the interpretation and justification of temporal memory operators in the wave equation. A generalization to the nonlinear longitudinal wave case is also given. This paper is an expanded version of \cite{holm2015four}.

\section{Background: Viscous wave equation}

The viscous wave equation in a homogeneous and isotropic medium, as assumed throughout this paper, is the same for both longitudinal and shear waves:
\begin{align}
\nabla^2 u -\dfrac 1{c_0^2}\frac{\partial^2 u}{\partial t^2} + \taus \dfrac{\partial}{\partial t}\nabla^2 u  = 0,
\label{eq:WaveViscous}
\end{align}
where $u(x,t)$ is longitudinal or shear displacement and $c_0$ is the zero frequency phase velocity. 

For longitudinal waves this velocity is $c_0^2 = (K+4 \mu/3)/\rho =(\lambda +2 \mu)/\rho$ where $K$ is the bulk modulus, $\lambda$ and $\mu$ are the first and second Lam\'{e}  parameters,  $\mu$ is also the shear modulus (often also denoted as $G$), and $\rho$ is the density. Fluids do not support shear forces, so $c_0^2 = K/\rho = \lambda/\rho$. This is a good approximation for tissue also as $K = \lambda+ 2\mu/3 \approx \lambda$ when $K \gg \mu$. The time constant is given as $\taus = \tau_{\sigma,L} = \eta_{L}/(\lambda +2 \mu) =  \eta_{L}/(\rho c_0^2)$, where $\eta_{L}$ is the longitudinal viscosity.

For shear waves one has $c_0^2 = \mu/\rho$ and $\taus = \tau_{\sigma,S} = \eta_{S}/\mu =  \eta_{S}/\rho c_0^2$, where $\eta_{S}$ is the shear viscosity. In tissue the Young's modulus is $E = \mu(3\lambda+2\mu)/(\lambda+\mu) \approx 3 \mu = 3 \rho c_0^2$.

This wave equation results in a power law attenuation $\alpha_k(\omega) \propto \omega^2$ and a constant phase velocity for $\omega \taus \ll 1$. For high relative frequencies,  $\omega \taus \gg 1$, both the attenuation and the phase velocity are  proportional to $\sqrt{\omega}$. 

This variation with frequency does not correspond well with longitudinal waves  in medical ultrasound at MHz frequencies which often exhibit near linear variation of attenuation with frequency. The corresponding dispersion is small, and often negligible \cite{Odonnell1981}. There is agreement that the main mechanism responsible for attenuation in ultrasound in tissue below about 15 MHz is absorption, as assumed in the viscous wave equation.

Shear waves in tissue elastography exhibit considerable dispersion in the 10-1000 Hz range which often deviates from the square law variation with frequency of Eq.  (\ref{eq:WaveViscous}) \cite{holm2010, Zhang2015shear}. In this case, it is more uncertain whether it is absorption or scattering which is the dominant mechanism and e.\ g.\ a multiple scattering model is found in \cite{Lambert2015bridging}. One difficulty in distinguishing between the two mechanisms  is that they both result in similar power law attenuation and dispersion. For that reason it may be that the following wave equations only are phenomenological without saying much about the mechanism. 

\subsection{Constitutive equations}

Eq.\ (\ref{eq:WaveViscous}) can be derived from a viscous constitutive equation, \cite{Royer00} (ch. 4.2.6):
\begin{align}
\sigma_{ij} = \lambda \delta_{ij} \boldmath{\epsilon} + 2 \mu \epsilon_{ij}  + (\eta_L - 2 \eta_S) \delta_{ij}\frac{\partial\epsilon}{\partial t} + 2 \eta_S \frac{\partial\epsilon_{ij}}{\partial t},
\label{eq:ConstTensor}
\end{align}
where $\epsilon = [\epsilon_{ij}(t)]$ and $\sigma = [\sigma_{ij}(t)]$ are the strain and stress tensors. The Kronecker delta is $\delta_{ij}=1$ if $i=j$ and zero otherwise. For a longitudinal wave, 
displacement and wave propagation are in the same direction, $i=j=1$. Then Eq.\ (\ref{eq:ConstTensor}) reduces to: 
\begin{align}
\sigma(t) = (\lambda + 2\mu) \left [\epsilon(t) +\tau_{\sigma,L} \frac{\partial\epsilon(t)}{\partial t} \right ],
\label{eq:ConstLong}
\end{align}

The shear wave constitutive equation when displacement is orthogonal to propagation ($j=1$, $i=2$), is:
\begin{align}
\sigma(t) = \mu \left[\epsilon(t) +\tau_{\sigma,S} \frac{\partial\epsilon(t)}{\partial t}\right].
\label{eq:ConstShear}
\end{align}
When combined with mass and momentum conservation laws, the longitudinal and shear wave versions of Eq.\ (\ref{eq:WaveViscous}) are found.

\subsection{Complex compressibility}

A different point of view can be developed by finding the ratio of strain and stress, the complex compressibility, in the frequency domain from Eqs.\ (\ref{eq:ConstLong}, \ref{eq:ConstShear}):
\begin{align}
\kappa(\omega) = \epsilon(\omega)/\sigma(\omega) 
= \frac{\kappa_0}{1+ i \omega \taus }= \kappa_0 - i \omega \frac{\tau_s \kappa_0}{1+ i \omega \taus }.
\label{Eq:dispersionKV}
\end{align}
Here the compressibility is $\kappa_0 = 1/(\lambda+2\mu)$ and $\kappa_0 = 1/\mu$ for the longitudinal and shear wave respectively \cite{holm2013deriving}.

\section{Frequency Domain: Power Law Attenuation (1)}
A wave equation with a convolution loss operator is the most versatile formulation \cite{Mainardi2010} (sect. 4.2.2):
\begin{align}
\nabla^2 u  - \frac{1}{c_0^2} \frac{\partial^2 u}{\partial t^2}  
+ \Phi(t) *  \nabla^2 u = 0,
\label{eq:Creep2}
\end{align}
where $\Phi(t) = [dG(t)/dt]/G_g$ is the rate of relaxation, $G(t)$ is the relaxation modulus, $G_g =G(0^+)$ is the glass modulus, and $E_e=G(\infty)$ is the equilibrium modulus. This modulus is $E_e = (\lambda + 2\mu)$ for the longitudinal wave, Eq.\ (\ref{eq:ConstLong}), and $E_e = \mu$ for the shear wave, Eq.\ (\ref{eq:ConstShear}). 

One particular form of the convolution operator is a power law memory in time. The operator is then characterized by an order parameter $\alpha$, where $m-1< \alpha <m$ and m is the smallest integer larger than the order:

\begin{align}
\Phi(t) * f(t) =\taus^\alpha  \frac{d^m f(t)}{dt^m}*\frac{1}{\Gamma(m-\alpha) t^{\alpha+1-m}}.
\end{align}
The scaling with the gamma function, $\Gamma(\cdot)$ ensures that its Fourier transform is also a power law \cite{Podlubny1999wholebook} (sect.\ 2.9):
\begin{align}
\Fourier\big(\frac{1}{\Gamma(m-\alpha) t^{\alpha+1-m}}\big) =(i\omega)^{\alpha-m}.
\end{align}
The power law characteristics makes this operator particularly interesting since that ensures that the solution of the wave equation also has  power law attenuation with an exponent in the most interesting range $\left( 0,2 \right]$.

Due to the similarity between the above Fourier transform and that of the integer order derivative, the time domain operator can also be written as a non-integer order derivative. Often this is called a fractional order derivative:
\begin{align}
\Phi(t) * f(t)= \taus^\alpha \frac{\partial^\alpha f(t)}{\partial t^\alpha}.
\end{align}

This leads to a fractional order wave equation \cite{Caputo1967, holm2010}:
\begin{align}
\nabla^2 u -\dfrac 1{c_0^2}\frac{\partial^2 u}{\partial t^2} + \taus^\alpha \dfrac{\partial^\alpha}{\partial t^\alpha}\nabla^2 u = 0.
\label{eq:KVWave}
\end{align}

The resulting attenuation, $\alpha_k$, and phase velocity, $c_p$, are:
\begin{align}
\alpha_k(\omega)  
   \begin{cases}
        = \alpha_0 \omega^{1+\alpha},         & (\omega \tau_\sigma)^{\alpha} \ll 1 \\
        \propto \omega^{1-\alpha/2},       & 1 \ll (\omega \tau_\sigma)^{\alpha} \\
    \end{cases}
	\label{eq:asymptote_KelvinVoigt}
\end{align}
\begin{align}
c_p(\omega) =
   \begin{cases}
	    c_0,				&  (\omega \tau_\sigma)^{\alpha} \ll 1 \\
    	\propto  \omega^{\alpha/2},       & 1 \ll  (\omega\tau_\sigma)^\alpha\\
   \end{cases}
\label{eq:asymptote_KelvinVoigtVelocity}
\end{align}

If $\alpha$ is in the range $0$ to $0.5$ this gives an attenuation which varies as $\omega^1$ to $\omega^{1.5}$ in the low-frequency (low $\omega \cdot \tau_\sigma$) regime. This can be used to model absorption encountered in medical ultrasound. This is the motivation for including such wave equations in a recent textbook for ultrasound \cite{szabo2014diagnostic} (ch.\ 4).

\section{Time Domain: Constitutive Equations (2)}

The constitutive stress-strain relation corresponding to the  wave equation of Eq.\ (\ref{eq:Creep2}) is also given by an operator:
\begin{align}
\sigma(t) =
 E_e \epsilon(t) + G(t)* \frac{\partial\epsilon(t)}{\partial t}.
\label{eq:convConst}
\end{align}
%
%

\begin{figure}[bt]
	\begin{center}
		\includegraphics[width=.35\columnwidth]{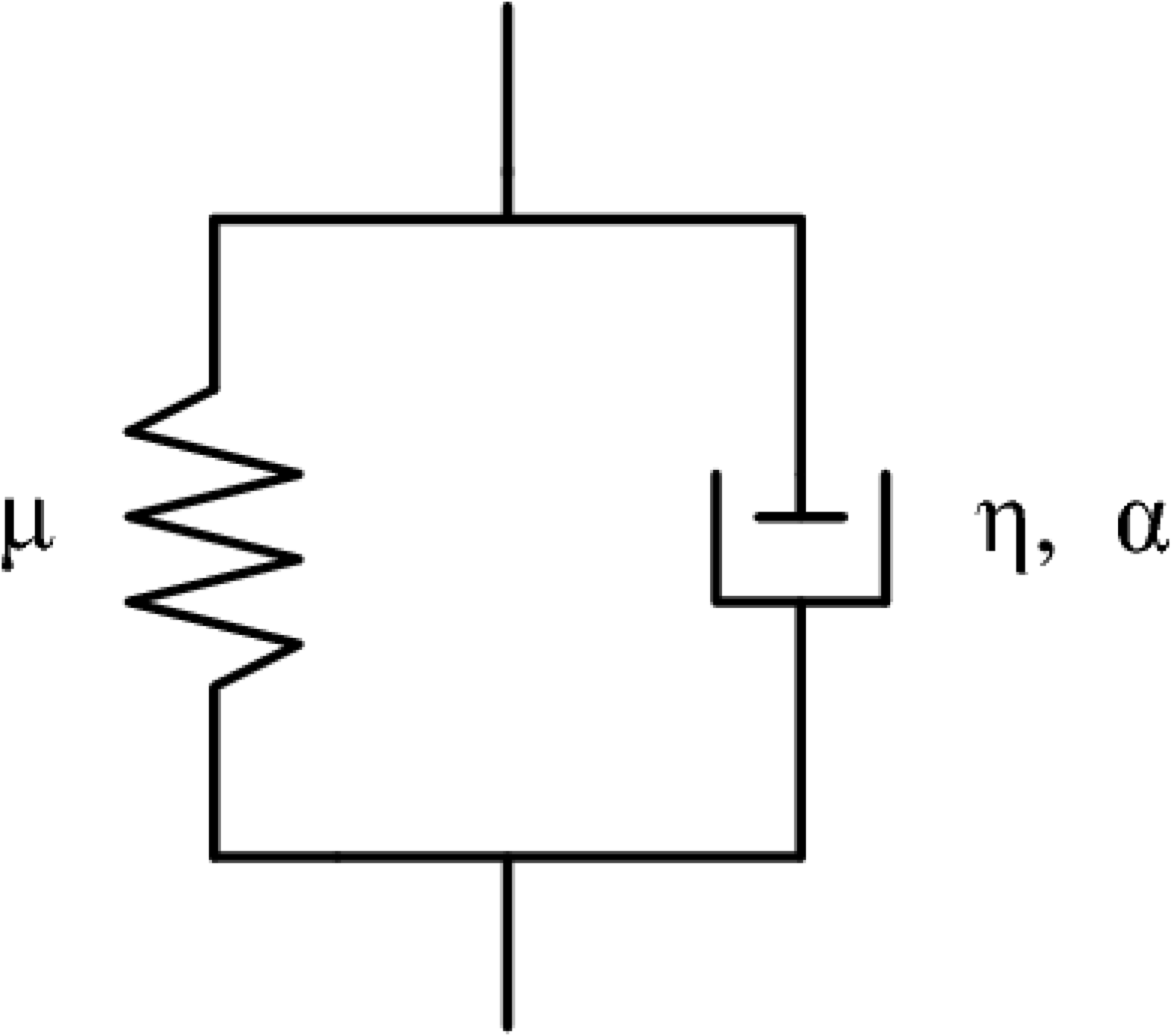}
	\end{center}
	\caption{Fractional Kelvin--Voigt model with spring characterized by shear modulus, $\mu$, and dashpot given by viscosity, $\eta$, and fractional order, $\alpha$, \cite{Zhang2015shear}.}
\label{fig:fractKV}
\end{figure}

When the fractional derivative is substituted in the above equation, the fractional Kelvin-Voigt mechanical model results, see Fig.\ (\ref{fig:fractKV}). Here it is expressed in the terminology of the shear wave of Eq.\ (\ref{eq:ConstShear}), but it could equally well have been expressed for longitudinal motion, Eq.\ (\ref{eq:ConstLong}):
\begin{align}
\sigma(t) =
 \mu \left[\epsilon(t) +\tau_{\sigma}^{\alpha} \frac{\partial^{\alpha}\epsilon(t)}{\partial t^{\alpha}}\right].
\label{eq:fractKVConst}
\end{align}

When combined with  mass and momentum conservation, the corresponding wave equation will be that of Eq.\ (\ref{eq:KVWave}).

The fractional Kelvin-Voigt model and its generalization, the fractional Zener model with fractional derivatives on both sides of the equal sign, fit well for biological materials such as brain, human root dentin, cranial bone, liver, arteries, breast, and hamstring muscle. They also fit other materials such as metals, doped corning glass, rubber, and polymers (see \cite{Nasholm2013Zener} and references therein). Most of these references apply to low frequency rheology, i.e.\ shear waves.

\section{A distribution of relaxation processes (3)}
Acoustic media such as salt water and air can be accurately described by a viscous part plus two physical relaxation mechanisms with parameters that are deduced from first principles. For salt water it is boric acid, $\mathrm{B(OH)_3}$, and magnesium sulfate, $\mathrm{MgSO_4}$, and for air it is nitrogen, $\mathrm{N_2}$, and oxygen, $\mathrm{O_2}$. 

Similar models of low order can be used to model arbitrary power law attenuation for more complex media over a limited frequency range \cite{Tabei2003}. In that case, the parameters of the model are not taken from physical processes, but rather selected to minimize a model fit error over the desired frequency range.

In general an N'th order relaxation process consists of N spring-damper terms,  each  with relaxation time $\tau_\nu = \eta_\nu/E_\nu$. This is the Maxwell-Wiechert model of Fig.\ ({\ref{fig:MW}) which can be expressed as a sum of compressibilities similar to Eq.\ (\ref{Eq:dispersionKV}):
\begin{align}
	\kappa(\omega) = \kappa_0-i\omega \sum_{\nu=1}^N \dfrac{\kappa_\nu\tau_\nu}{1 + i\omega\tau_\nu},
\label{eq:compress}
\end{align}
where the compressibilities are $\kappa_\nu= 1/E_\nu$.
\begin{figure}[bt]
	\begin{center}
		\includegraphics[width=.65\columnwidth]{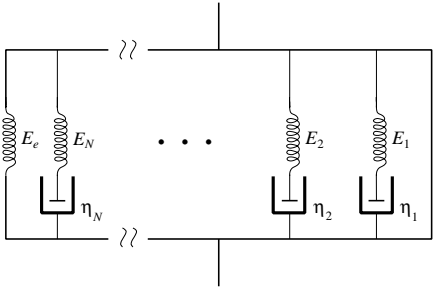}
	\end{center}
	\caption{Maxwell--Wiechert model, \cite{Nasholm2013Zener}.}
\label{fig:MW}
\end{figure}

The fractional Zener model can be interpreted as a distribution of an infinite number of relaxation models, $N \rightarrow \infty$ \cite{mainardi1994fractional}. Thus, the medium can be understood as being so complex that it takes an infinite number of processes to describe it, rather than two or three as for salt water and air. The particular weighting, $\kappa_\nu(\Omega_\nu)$, $\Omega_\nu = 1/\tau_\nu$, of each relaxation process follows a Mittag-Leffler function as a function of relaxation frequency, $\Omega$. This function has a power law tail and thus indicates fractal properties in the medium, \cite{Nasholm2011}. This is a property that probably can be exploited more.


\section{Higher order derivatives (4)}
The Maxwell-Wiechert 
model of Fig.\ ({\ref{fig:MW}) is  equivalent to a linear differential equation of order N, \cite{tschoegl1989phenomenological} (sect. 3.7), and \cite{Mainardi2010} (sect. 2.4), and this gives an alternative to Eq.\ (\ref{eq:fractKVConst}):
\begin{align}
	\sum_{n=0}^{N} p_n \frac{\partial^{n}\sigma(t)}{\partial t^{n}}  = \sum_{n=0}^{N} q_n \frac{\partial^{n}\epsilon(t)}{\partial t^{n}}.
\label{Eq:highOrder}
\end{align}
where $p_n$ and $q_n$ are constants. A fractional derivative expressed with higher order derivatives can also be found in \cite{oldham1974fractional} (sect. 3.5), from a mathematical point of view.

When $N \rightarrow \infty$ and the coefficients are chosen using similar reasoning as in the previous section, this constitutive equation can describe a fractional one like Eq.\ (\ref{eq:fractKVConst}). Higher order derivatives are conceptually simple and may offer a more intutive explanation than one using fractional derivatives.

However, the advantage of the fractional description is parsimony.
As an example Eq.\ (\ref{eq:fractKVConst}) requires only $E_e, \taus$ and $\alpha$ while  Eq.\ (\ref{Eq:highOrder}) in principle requires an infinite number of parameters. In practice this number can probably be truncated, but nevertheless it will always exceed three.

\section{Fractional Westervelt Equations}

Shear waves will experience a symmetric nonlinearity \cite{catheline2003observation} due to a third order term in Eq.\ (\ref{eq:ConstShear}), while here we will only cover longitudinal waves where the medium behaves differently during  compression and  rarefaction phases.

The Westervelt equation assumes weak nonlinearity and builds on two independent 
loss mechanisms. The first is due to viscous losses as discussed before for  Eq.\ (\ref{eq:ConstLong}).
The second is due to thermal conduction between the higher temperature compressed parts of the wave and lower temperature parts during rarefaction, and only plays a role in gases. 
The thermal time constant is:
\begin{equation}
\tau_{th} = \frac{\kappa}{\rho_0 c_0^2}\left(\frac{1}{c_v}-\frac{1}{c_p}\right)
\end{equation}
where $c_v$ and $c_p$ are the specific heat capacities for constant volume and constant pressure respectively. 

In the derivation of the fractional Westervelt equation, the thermal loss effect can also be formulated with fractional derivatives by generalizing the Fourier heat law \cite{prieur2011, prieur2012, holm2013deriving}. In its standard form it is  $\mathbf{q} = -\kappa_{th}{\nabla} T$ where $\mathbf{q}$ is heat flow, $T$ is temperature, and  $\kappa_{th}$ is the thermal conductivity. It predicts a counterintuitive infinite heat propagation velocity \cite{zhang2014}, which can be made finite with a fractional heat law:
\begin{equation}
\mathbf{q}(t) = - \kappa_{th} \tau_{th}^{1-\gamma}   \frac{\partial^{1-\gamma}}{\partial t^{1-\gamma}}  {\nabla} T(t),
\end{equation}
where $\tau_{th}$ is a thermal relaxation time. The fractional order is $\gamma$ where $\gamma=1$ leads to the Fourier heat law.

The fractional Westervelt equation of the first form has independent fractional orders in the two loss terms:
\begin{equation}
\nabla^2 p - \frac{1}{c_0^2}\frac{\partial^2 p}{\partial t^2} 
+\tau_{\sigma,L}^\alpha\frac{\partial^\alpha}{\partial t^\alpha} \nabla^2 p 
+ \tau_{th}^{2-\gamma}\frac{\partial^{2-\gamma}}{\partial t^{2-\gamma}}  \nabla^2 p
= -\frac{\beta_{NL}}{\rho_0 c_0^4}\frac{\partial^2 p^2}{\partial t^2}.
\end{equation}

Here $p$ is pressure 
and $\gamma$ is the fractional order due to thermal effects. 
The  nonlinearity parameter is $\beta_{NL} = 1 + B/2A$. Its two terms show the origin of nonlinearity in two effects: quadratic  nonlinearity ($B/2A$) in  Eq.\ (\ref{eq:ConstLong})  and nonlinearity due to convection stemming from nonlinearity in the Navier-Stokes equation (the $1$ in the expression for $\beta_{NL}$). 

In the second form, the orders are coupled, $\gamma = 2-\alpha$, and the time constants are combined, $\tau = \tau_{\sigma,L} +  \tau_{th}$:
\begin{equation}
\nabla^2 p - \frac{1}{c_0^2}\frac{\partial^2 p}{\partial t^2} 
+ \tau^{\alpha} \frac{\partial^\alpha}{\partial t^\alpha} \nabla^2 p 
= -\frac{\beta_{NL}}{\rho_0 c_0^4}\frac{\partial^2 p^2}{\partial t^2}
\label{eq:Westervelt2}
\end{equation}


\subsection{Low frequency approximations}

When frequencies are low, i.e. $\omega  \tau^{\alpha} \ll 1$, the losses are small and the lossless approximation ${c_0^2} \nabla^2 p \approx \partial^2 p/\partial t^2$ applies. Inserting that in the loss term of Eq.\ (\ref{eq:Westervelt2}) gives:
\begin{equation}
\nabla^2 p - \frac{1}{c_0^2}\frac{\partial^2 p}{\partial t^2} 
+\frac{\tau^{\alpha}}{c_0^2} \frac{\partial^{\alpha+2}p}{\partial t^{\alpha+2}}
= -\frac{\beta_{NL}}{\rho_0 c_0^4}\frac{\partial^2 p^2}{\partial t^2}
\label{eq:Westervelt2time}
\end{equation}

A common form of the Westervelt equation can be recognized for $\alpha=1$. 

Computation is often faster with a spatial domain  operator. This is because of the smaller amount of data over the finite support in space compared to that of the infinite time history.  The transformation to space is somewhat involved and can be found in the derivation leading up to Eq.\ (27) in \cite{holm2014comparison}: 
\begin{align}
&\nabla^2 p  - \frac{1}{c_0^2} \frac{\partial^2 p}{\partial t^2}
- 2  \alpha_0 c_0^{\alpha} \frac{\partial}{\partial t} (-\nabla^2)^{(\alpha+1)/2} p\notag\\
&-2\alpha_0c_0^{\alpha+1} \tan(\pi \frac{\alpha+1}{2}) (-\nabla^2)^{\alpha/2+1} p = -\frac{\beta_{NL}}{\rho_0 c_0^4}\frac{\partial^2 p^2}{\partial t^2},
\label{eq:fractLaplacianCausal2}
\end{align}
where 
$\alpha_0 =  0.5 c_0^{-1} \sin(\pi\alpha/2) \tau_{\sigma}^\alpha$ in Eq.\ (\ref{eq:asymptote_KelvinVoigt}).
The non-causal, nonlinear version 
(no second loss term) was  shown in \cite{chen2002fractional}, see also \cite{Chen04}. The causal, linear version ($\beta_{NL}=0$) was first shown in \cite{Treeby2010sectionIIB} and is also used in the k-Wave MATLAB toolbox \cite{treeby2014modeling}. 

\section{Conclusion}
One way to model arbitrary power law attenuation is via temporal memory operators. In order to further the understanding of this point of view, four interpretations have been given here with the hope of increasing of the acceptance of these models in ultrasonics and elastrography. 

Further work involves deriving the constitutive relations from a specific physical mechanism as started in \cite {Pandey2015fractional}.






\end{document}